\documentclass[aps,prl,reprint,groupedaddress,nofootinbib,floatfix,longbibliography,eprint]{revtex4-2}

\usepackage{siunitx} 
\usepackage{amsmath} 
\usepackage{amsthm} 
\usepackage{amssymb}	
\usepackage{capt-of}
\usepackage{graphicx} 
\usepackage{exscale,fix-cm} 
\usepackage{enumerate}

\usepackage{bbm}

\usepackage[colorlinks=true,linkcolor=blue,citecolor=blue,urlcolor=blue]{hyperref}

\renewcommand{\v}[1]{\ensuremath{\mathbf{#1}}} 
\newcommand{\ket}[1]{| #1 \rangle } 
 
\let\baraccent=\= 
\renewcommand{\=}[1]{\stackrel{#1}{=}} 

\def\C{\mathcal{C}}

\def\H{\mathcal{H}}

\def\Tr{\text{Tr}}

\usepackage[T1]{fontenc}
\usepackage{bm}
\usepackage{xcolor}

\usepackage{needspace}
\usepackage{tikz}
\usetikzlibrary{arrows.meta}

\definecolor{WedgeBlue}{RGB}{40,91,155}
\definecolor{WedgeGreen}{RGB}{32,125,108}
\definecolor{BraneRed}{RGB}{173,49,72}
\definecolor{RevisionBlue}{RGB}{0,65,190}

\def\E{\mathcal{E}}

\def\Hbb{\mathbb{H}}

\newcommand{\Area}{\text{Area}}
\newcommand{\wtS}{\widetilde{S}}

\begin{document}

\title{Refined R\'enyi Entropy from Replica-Geometry Tensor Networks}

\author{Qiang Wen$^{a}$}
\email{wenqiang@seu.edu.cn}

\author{Haocheng Zhong$^{a,1}$}

\author{Yiwei Zhong$^{a,1}$}

\affiliation{$^a$School of Physics and Shing-Tung Yau Center, Southeast University, Nanjing 210096, China}

\begin{abstract}
Holographic tensor networks are a central route to understanding the mechanism behind the holographic duality, but standard fixed-graph constructions with maximally entangled links face a structural obstacle. Their leading entanglement spectra are flat, so their R\'enyi entropies do not depend on the replica index. This is in sharp contrast with holographic conformal field theories, where a tensionful cosmic brane makes the refined R\'enyi entropy $n$-dependent. Capturing this replica dependence is therefore an important test of tensor-network descriptions of holographic entanglement. We construct an explicit family of replica-dependent geodesic graphs that reproduces this behavior while keeping the links maximally entangled. For each replica index $n$, geodesics are emitted from the replicated boundary with the same local, isotropic intersection measure. They perfectly tessellate the selected replica geometry, so the $n$-dependence of their intersection counts comes from the geometry. Tensor-network models on these graphs reproduce the cosmic-brane prescription for refined R\'enyi entropy by counting crossing threads. Their cross-copy endpoint density also agrees, in CFT$_2$, with a boundary reconstruction within the additive linear combination proposal, providing a check beyond the total entropy. In higher dimensions, we further obtain an analytic large-separation decay law for this density.
\end{abstract}

\maketitle
\footnotetext[1]{These authors contributed equally to this work.}
\flushbottom

\textcolor{blue}{\emph{Introduction.}} The Anti-de Sitter/Conformal Field Theory (AdS/CFT) correspondence, together with its gauge/gravity generalizations, is our principal window into quantum gravity~\cite{Maldacena:1997re}. Among its most important achievements is a nonperturbative definition of quantum gravity in asymptotically AdS spacetimes, from which black-hole thermodynamics and microstate counting~\cite{Strominger:1996sh,Callan:1996dv,Horowitz:1996fn,Maldacena:1996ky,Strominger:1997eq}, strongly coupled gauge dynamics, and the geometric encoding of entanglement all follow. A vast literature has tested the holographic dictionary by computing the same physical quantity on both sides and finding agreement~\cite{Gubser:1998bc,Witten:1998qj,Freedman:1998aa,Witten:1998zw,Klebanov:1999tb,Ryu:2006bv,Ryu:2006ef,Son:2002sd,Henningson:1998ey,Maldacena:2015wa,Dutta:2019gen,Headrick:2010renyi,Wen:2021bal}. These successes, however, establish the dictionary rather than explain its mechanism. To understand this mechanism, an important line of research has been the construction of tensor-network models of holographic duality. Entanglement renormalization~\cite{Swingle:2009cg} explains how a bulk radial direction emerges from boundary entanglement. The Harlow--Pastawski--Preskill--Yoshida (HaPPY) code~\cite{Pastawski:2015qua} reproduces the Ryu--Takayanagi (RT) formula~\cite{Ryu:2006bv,Ryu:2006ef}, bulk locality, and the quantum error-correcting structure of the holographic map. Random tensor networks~\cite{Hayden:2016cfa} satisfy the RT-like formula and its generalizations at large bond dimension, as typical entanglement properties of random boundary states. See Refs.~\cite{Apel:2022,Qi:2017,Chandra:2023} for further developments along this direction, where random tensor networks are combined with HaPPY-type quantum error correction. Together, these models capture, one by one, the geometric encoding of entanglement, subregion duality, bulk reconstruction, and the resilience of bulk physics to boundary erasure. 

Yet the fixed-graph models with maximally entangled links share a structural limitation. The entanglement is computed by minimal cuts whose size is independent of the replica index, so the leading-order R\'enyi entropies are $n$-independent, as for a reduced density matrix with a flat spectrum. In AdS/CFT, the R\'enyi entropies are computed on the gravity side using the replica trick~\cite{Lewkowycz:2013nqa,Dong:2016fnf,Dong:2023modified} and have a nontrivial $n$-dependence, in agreement with analytic CFT computations~\cite{Calabrese:2004eu,Hung:2011nu}. This differs sharply from the flat-spectrum networks. This tension shows that these fixed-graph networks capture the area law but not the full entanglement spectrum of holographic states. Resolving the flat-spectrum problem is therefore essential for tensor networks to become faithful models of quantum gravity.

The \(n\)-dependence of holographic R\'enyi entropies is by now well understood. Applying the replica trick in the boundary CFT produces a corresponding replicated bulk geometry, and at leading order the gravitational path integral is dominated by a saddle point described by a classical metric on the replica manifold. This saddle depends on the replica index \(n\), so the cosmic brane that computes the refined R\'enyi entropy moves with \(n\)~\cite{Lewkowycz:2013nqa,Dong:2016fnf}. A holographic tensor network consists of the tensors (or quantum states) placed at each site, and the graph on which they are contracted. The latter is supposed to model the bulk background geometry. In these fixed-graph networks, however, the graph is prescribed rather than determined by the replica saddle. Consequently, when the entanglement or R\'enyi entropy is computed by counting minimal cuts, neither the number of cut links nor the path of the minimal cut changes with \(n\). If the gravitational replica manifold is \(n\)-dependent, the graph that models the background geometry should exhibit the same \(n\)-dependence. However, the fixed-graph models described above do not incorporate this geometric backreaction. Related works have drawn an analogy between tensor networks and fixed-area states of AdS/CFT~\cite{Dong:2018seb,Akers:2018fow}. In this picture, a general semiclassical state whose bulk dual is a smooth classical geometry with small quantum fluctuations can be viewed as a superposition of fixed-area states, and in the R\'enyi path integral different area saddles dominate at different \(n\), producing the \(n\)-dependence. Alternatively, one can keep the graph fixed but replace the maximally entangled links by non-maximally entangled link states. This directly modifies the entanglement spectrum and can generate \(n\)-dependent R\'enyi entropies without superposing different area sectors~\cite{Dong:2021neg,Cheng:2022}. See Refs.~\cite{Dong:2023,Akers:2021,Ling:2019,Bao:2019,Held:2024bgs,Dong:2025crd,Dong:2019piw,Cao:2023mzo,Akers:2024ixq,Han:2017uco,Dong:2024gud,Bhattacharyya:2016hbx,Penington:2022dhr} for other approaches to non-flat spectra in holographic tensor-network models. 

Interesting complementary tensor-network approaches based on boundary CFT data have recently been developed~\cite{Chen:2024unp,Bao:2024ixc,Geng:2025efs} (see also Ref.~\cite{Han:2017uco}). In these tensor networks the links carry labels summed with weights from CFT datas. At large central charge, the saddle labels describe geodesic lengths, and weights reproduce the Einstein--Hilbert action. The CFT tensors are quasi-perfect isometries that retain the CFT entanglement spectrum, with non-flat R\'enyi entropies arising from summing over geometries. These constructions currently focus on AdS$_3$/CFT$_2$, special CFTs or ensembles with large central charge, and the $n$-dependent cosmic-brane saddle is not made explicit.

Our aim is to construct an explicit family of tensor-network graphs that follows the changing replica geometry while keeping the local counting rule and link states fixed. The fixed-area-state picture explains how different area saddles dominate at different \(n\)~\cite{Dong:2018seb,Akers:2018fow}. Ref.~\cite{Akers:2018fow} also suggested superposing networks with different graph structures. Building on the geodesic tessellations and PEE tensor networks of Refs.~\cite{Wen:2025gui,Basu:2026hbg,Gu:2026rfo}, we make the replica-dependent graph explicit and compute its cut entropy. Beyond the total entropy, we determine the cross-copy endpoint distribution, verify it against a CFT$_2$ reconstruction within the additive linear combination proposal, and extract its large-separation decay in higher dimensions.

\textcolor{blue}{\emph{Graphs as perfect tessellations of geometry.}} For homogeneous Riemannian manifolds such as pure AdS space and Euclidean flat space, graphs that perfectly tessellate the geometry can be found in integral geometry~\cite{Santalo,Helgason,Czech:2015qta}, which is formed by the superposition of all geodesics in kinematic space. More explicitly, kinematic space \(\mathcal K\) is the space of all unoriented geodesics in the bulk. Each point \(\gamma\in\mathcal K\) labels one geodesic, and the Crofton form endows \(\mathcal K\) with a natural measure \(d\mu_{\mathcal K}\). The Crofton formula then states that the area of any bulk hypersurface \(\Sigma\) is obtained by counting its intersections with the geodesics, \(\Area(\Sigma)=\frac{1}{\Omega_{d-2}}\int_{\mathcal K}\#(\gamma\cap\Sigma)\,d\mu_{\mathcal K}\), where \(\#(\gamma\cap\Sigma)\) counts the number of transverse intersections and \(\Omega_{d-2}\) is the area of the unit \((d-2)\)-sphere~\cite{Czech:2015qta,Czech:2016xec,Santalo,Lin:2023rxc,Lin:2024fze}. Such continuous geodesic graphs therefore give a perfect tessellation of the background geometry. The tessellation was then generalized to subregions of a non-homogeneous manifold, with the geodesics of kinematic space replaced by geodesic chords shot from the boundary of the subregion with a suitable measure~\cite{Gu:2026rfo,Basu:2026hbg}. For example, the exterior of a Schwarzschild black hole can be perfectly covered by geodesics emitted from a boundary spherical shell, and the exterior of a Ba\~nados--Teitelboim--Zanelli (BTZ) black hole can be perfectly tessellated by geodesics emitted from the AdS boundary~\cite{Basu:2026vsq}.

Previous tensor network models compute the entanglement entropy of a boundary region by counting the minimal number of bonds cut by a surface in a fixed graph. This is not yet the area of an extremal surface in AdS/CFT, because the minimal graph cut and the area of the minimal surface are related only indirectly. Refs.~\cite{Wen:2025gui,Basu:2026hbg,Gu:2026rfo} defined the partial-entanglement-entropy (PEE) tensor-network models (see Appendix~\ref{app:pee-models} for a brief review) on geodesic graphs that perfectly tessellate the geometry. In these models, the number of bonds in the minimal cut is, by the Crofton formula, exactly the area of the minimal surface, so the RT formula is reproduced exactly.

Following this route, we construct a geodesic network that perfectly tessellates the replica AdS manifold for each \(n\), define a tensor-network model on it, and then compute the refined R\'enyi entropy directly in this network and compare it with the cosmic brane prescription. Throughout, we consider Einstein gravity and restrict to the time-reflection-symmetric slice of a configuration, and set the AdS curvature radius $L_{\rm AdS}=1$ unless explicitly restored. The point is that $\wtS_n(A)$ is itself a von Neumann entropy of the normalized power state $\rho_{A,n}=\rho_A^n/\Tr\rho_A^n$, so we use the same thread-counting rule to compute it in the replica PEE tensor network, which reproduces $\wtS_n(A)=\Area(\mathcal C_n)/4G_{\rm N}$, with the $n$-dependence carried by the geometry.

\textcolor{blue}{\emph{Refined R\'enyi entropy and the cosmic-brane prescription.}} Let $A$ be a spatial region and $\rho_A$ its reduced density matrix, normalized by $\Tr\rho_A=1$. For real $n>0$, define the partition function $Z_n=\Tr\rho_A^n$ and the R\'enyi entropy $S_n(A)=\log Z_n/(1-n)$ with the von Neumann entropy recovered as $n\to1$. We further define the normalized power state as $\rho_{A,n}=\rho_A^n/Z_n$ such that the refined R\'enyi entropy obeys \cite{Dong:2016fnf,Nakaguchi:2016zqi}
\begin{equation}
	\begin{aligned}
		 \wtS_n(A)
		&\equiv n^2\partial_n\!\left(\frac{n-1}{n}S_n(A)\right)\\&=-\Tr\!\left(\rho_{A,n}\log\rho_{A,n}\right)=S(\rho_{A,n}),
	\end{aligned}
 \label{eq:refined-power}
\end{equation}
which is just the von Neumann entropy of $\rho_{A,n}$. The path-integral replica construction below starts at integer $n$ and results at noninteger $n$ refer to its continuation along the chosen replica-symmetric saddle branch.

For integer $n\geq2$, the replica construction computes $Z_n$ by cyclically sewing $n$ copies of the boundary path integral along $A$, which has a gravity dual where $Z_n$ is computed at the leading order by the classical bulk gravity solution $\mathcal B_n$ preserving the cyclic permutation symmetry $\mathbb Z_n$ \cite{Lewkowycz:2013nqa,Dong:2016fnf,Calabrese:2004eu,Headrick:2010renyi}. The quotient $\widehat{\mathcal B}_n=\mathcal B_n/\mathbb Z_n$ contains a codimension-two cosmic-brane locus $\mathcal{C}_{n}$ with tension $T_n=\frac{n-1}{4nG_{\rm N}}$. In the replicated geometry $\mathcal B_n$, $\mathcal{C}_{n}$ is the fixed-point set of the $\mathbb Z_n$ replica symmetry. In the quotient, it is a tensionful extremal brane whose location follows from extremizing the gravitational action. The cosmic-brane prescription for the refined R\'enyi entropy at leading order is therefore an analogue of the RT formula \cite{Lewkowycz:2013nqa,Dong:2016fnf},
\begin{equation}
	\wtS_n(A)=\frac{\Area(\mathcal{C}_{n})}{4G_{\rm N}}.
	\label{eq:cosmic-law}
\end{equation}

More explicitly, let us consider a boundary spherical region $A=\{x\in\mathbb R^{d-1}:|x|<R\}$ in the vacuum of the boundary CFT. The Casini--Huerta--Myers (CHM) map identifies its reduced state with a thermal state on hyperbolic space~\cite{Casini:2011kv}. On a static time slice, the CHM map reduces to 
\begin{equation}
	x=R\tanh\frac{\eta}{2}\,\hat x,
	\label{eq:chm-spatial-map}
\end{equation}
where $\hat x\in S^{d-2}$ and $\eta$ is the hyperbolic radial coordinate, and $dx\cdot dx=\Lambda(x)^2[d\eta^2+\sinh^2\eta\,d\Omega_{d-2}^2]=\Lambda(x)^2dH_{d-1}^2$ with $\Lambda(x)=\frac{R^2-|x|^2}{2R}$. The map \eqref{eq:chm-spatial-map} is independent of \(n\). The replica index enters only through the inverse temperature $\beta_n=2\pi n R$ of the thermal state. The bulk extension of the CHM map is a spatial isometry from the spatial wedge region $W_n$ between $A$ and the locus $\mathcal C_n$ on the time-reflection-symmetric slice of the selected backreacted saddle to the exterior of a hyperbolic black hole described by the following metric~\cite{Hung:2011nu,Dong:2016fnf,Agon:2025fluids},
\begin{equation}\label{eq:spatial hyperbolic}
	\begin{aligned}
		h_{\Hbb}
		&=\frac{dr^2}{f_n(r)}+r^2dH_{d-1}^2,
		\quad r\ge r_n,\\
		f_n(r)&=r^2-1-\frac{\mu_n}{r^{d-2}},\quad\mu_n=r_n^d-r_n^{d-2},\\
		r_n&=\frac{1+\sqrt{1+d(d-2)n^2}}{dn},
	\end{aligned}
\end{equation}
where $d$ is the boundary spacetime dimension, $dH_{d-1}^2$ is the unit hyperbolic metric, and regularity fixes $4\pi/f'_n(r_n)=2\pi n$. The bulk CHM map then identifies $\rho_{A,n}$ as a thermal state at inverse temperature $\beta_n=2\pi nR$
and maps the extremal cosmic brane $\mathcal C_n$ to the extremal horizon cross-section $\E_n$ (see Appendix~\ref{app:chm-purification}). Since the bulk CHM is an isometry,
\begin{equation}
	\Area(\C_n)=\Area(\E_n)=r_n^{d-1}V_{\mathbb H},
	\label{eq:brane-regulated-area}
\end{equation}
where $V_{\mathbb H}$ is the regulated unit hyperbolic volume at fixed hyperbolic cutoff~\cite{Casini:2011kv,Hung:2011nu}.

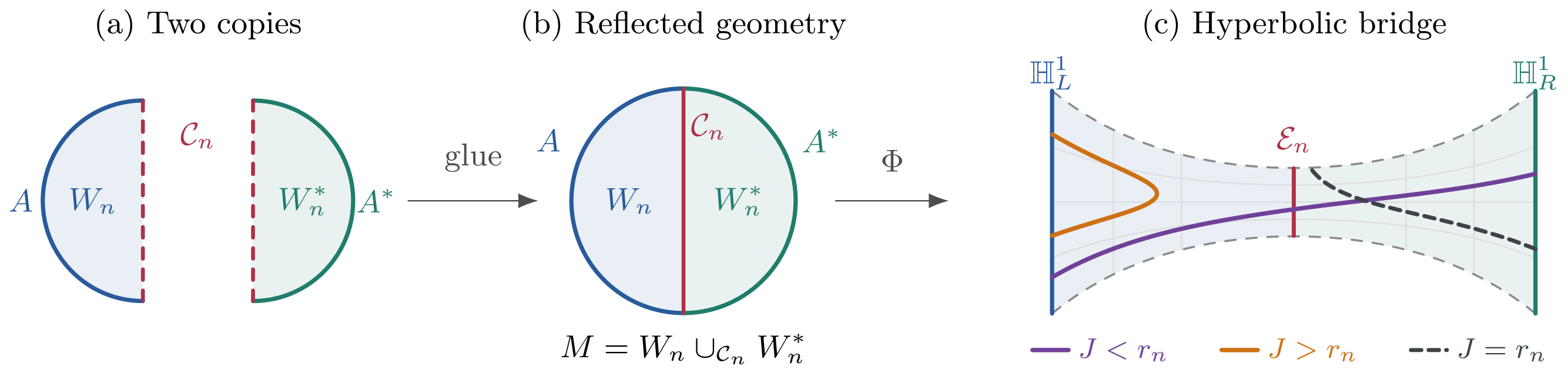
\begin{figure*}[t]
\centering
\definecolor{GeoCross}{RGB}{112,65,153}
\definecolor{GeoReturn}{RGB}{206,112,20}
\definecolor{GeoCritical}{RGB}{62,67,73}
\newsavebox{\combinedfigurebox}
\sbox{\combinedfigurebox}{\resizebox{0.98\textwidth}{!}{%
\begin{tikzpicture}[x=1cm,y=1cm,font=\small,line cap=round,line join=round]
\node at (1.55,1.75) {(a) Two copies};
\node at (6.4,1.75) {(b) Reflected geometry};
\node at (12.5,1.75) {(c) Hyperbolic bridge};
\fill[WedgeBlue!10] (1,1) arc[start angle=90,end angle=270,radius=1] -- cycle;
\fill[WedgeGreen!10] (2.1,1) arc[start angle=90,end angle=-90,radius=1] -- cycle;
\draw[WedgeBlue,line width=1.2pt] (1,1) arc[start angle=90,end angle=270,radius=1];
\draw[WedgeGreen,line width=1.2pt] (2.1,1) arc[start angle=90,end angle=-90,radius=1];
\draw[BraneRed,dashed,line width=1.1pt] (1,-1)--(1,1);
\draw[BraneRed,dashed,line width=1.1pt] (2.1,-1)--(2.1,1);
\node[text=WedgeBlue] at (.5,0) {$W_n$};
\node[text=WedgeGreen] at (2.6,0) {$W_n^*$};
\node[text=WedgeBlue] at (-.22,0) {$A$};
\node[text=WedgeGreen] at (3.33,0) {$A^*$};
\node[text=BraneRed] at (1.55,.65) {$\mathcal C_n$};
\draw[-{Latex[length=2mm]},black!70] (3.65,0)--(4.95,0) node[midway,above=5pt] {glue};
\fill[WedgeBlue!10] (6.4,1.12) arc[start angle=90,end angle=270,radius=1.12]--cycle;
\fill[WedgeGreen!10] (6.4,1.12) arc[start angle=90,end angle=-90,radius=1.12]--cycle;
\draw[WedgeBlue,line width=1.2pt] (6.4,1.12) arc[start angle=90,end angle=270,radius=1.12];
\draw[WedgeGreen,line width=1.2pt] (6.4,1.12) arc[start angle=90,end angle=-90,radius=1.12];
\draw[BraneRed,line width=1.1pt] (6.4,-1.12)--(6.4,1.12);
\node[text=WedgeBlue] at (5.86,0) {$W_n$};
\node[text=WedgeGreen] at (6.95,0) {$W_n^*$};
\node[text=WedgeBlue] at (5.05,.6) {$A$};
\node[text=WedgeGreen] at (7.78,.6) {$A^*$};
\node[text=BraneRed] at (6.65,.74) {$\mathcal C_n$};
\node at (6.4,-1.48) {$M=W_n\cup_{\mathcal C_n}W_n^*$};
\draw[-{Latex[length=2mm]},black!70] (7.92,0)--(9.05,0) node[midway,above=5pt] {$\Phi$};
\node[inner sep=0pt,anchor=center] at (12.5,-.08) {\begin{tikzpicture}[
			x=.9cm,y=.62cm,
			font=\small,text=black,
			line cap=round,line join=round,
			geodesic/.style={line width=1.25pt},
			declare function={
				geowidth(\uu)=0.55*cosh(\uu);
				crossheight(\uu)=-0.30+ln(0.75*tanh(\uu)+sqrt(1+0.5625*tanh(\uu)^2));
				returnradial(\vv)=-0.5*ln((1+0.7806247498*cosh(\vv-0.22))/(1-0.7806247498*cosh(\vv-0.22)));
				critc=0.6;
				criticalradial(\vv)=0.5*ln((1+tanh(1.85)*exp(-\vv-critc))/(1-tanh(1.85)*exp(-\vv-critc)));
				criticaly(\vv)=geowidth(criticalradial(\vv))*0.7*\vv;
			}
			]
			\fill[WedgeBlue!10]
			plot[domain=-1.85:0,variable=\uu,samples=45] ({1.45*\uu},{geowidth(\uu)})
			-- plot[domain=0:-1.85,variable=\uu,samples=45] ({1.45*\uu},{-geowidth(\uu)}) -- cycle;
			\fill[WedgeGreen!10]
			plot[domain=0:1.85,variable=\uu,samples=45] ({1.45*\uu},{geowidth(\uu)})
			-- plot[domain=1.85:0,variable=\uu,samples=45] ({1.45*\uu},{-geowidth(\uu)}) -- cycle;
			\foreach \ss in {-1.38,-.86,.86,1.38}{
				\draw[black!13,line width=.4pt]
				({1.45*\ss},{-geowidth(\ss)}) -- ({1.45*\ss},{geowidth(\ss)});
			}
			\foreach \tt in {-.5,0,.5}{
				\draw[black!13,line width=.4pt]
				plot[domain=-1.85:1.85,variable=\uu,samples=65]
				({1.45*\uu},{\tt*geowidth(\uu)});
			}
			\foreach \tt in {-1,1}{
				\draw[black!45,dashed,line width=.6pt]
				plot[domain=-1.85:1.85,variable=\uu,samples=65]
				({1.45*\uu},{\tt*geowidth(\uu)});
			}
			\draw[WedgeBlue,line width=1.2pt]
			(-2.6825,{-geowidth(1.85)}) -- (-2.6825,{geowidth(1.85)});
			\draw[WedgeGreen,line width=1.2pt]
			(2.6825,{-geowidth(1.85)}) -- (2.6825,{geowidth(1.85)});
			\draw[BraneRed,line width=1.15pt]
			(0,{-criticaly(1.4)}) -- (0,{criticaly(1.4)});
			\node[text=WedgeBlue] at (-2.6825,2.08) {$\mathbb H_L^1$};
			\node[text=WedgeGreen] at (2.6825,2.08) {$\mathbb H_R^1$};
			\node[text=BraneRed] at (0,1.) {$\E_n$};
			
			\draw[GeoCross,geodesic]
			plot[domain=-1.85:1.85,variable=\uu,samples=110]
			({1.45*\uu},{geowidth(\uu)*0.7*crossheight(\uu)});
			\draw[GeoReturn,geodesic]
			plot[domain=-.4305985853:.8705985853,variable=\vv,samples=100]
			({1.45*returnradial(\vv)},{geowidth(returnradial(\vv))*0.7*\vv});
			\draw[GeoCritical,densely dashed,line width=1.15pt]
			plot[domain=-critc:1.4,variable=\vv,samples=110]
			({1.45*criticalradial(\vv)},{criticaly(\vv)});
			
			\draw[GeoCross,line width=1.25pt] (-2.90,-2.38) -- (-2.50,-2.38);
			\node[anchor=west,font=\footnotesize,text=GeoCross] at (-2.40,-2.38) {$J<r_n$};
			\draw[GeoReturn,line width=1.25pt] (-.80,-2.38) -- (-.40,-2.38);
			\node[anchor=west,font=\footnotesize,text=GeoReturn] at (-.30,-2.38) {$J>r_n$};
			\draw[GeoCritical,densely dashed,line width=1.15pt] (1.30,-2.38) -- (1.70,-2.38);
			\node[anchor=west,font=\footnotesize,text=GeoCritical] at (1.80,-2.38) {$J=r_n$};
		\end{tikzpicture}};
\end{tikzpicture}}}
\typeout{PRL-FIGURE-WIDTH=\the\wd\combinedfigurebox}
\typeout{PRL-FIGURE-HEIGHT=\the\dimexpr\ht\combinedfigurebox+\dp\combinedfigurebox\relax}
\usebox{\combinedfigurebox}
\caption{Reflected geometry and crossing threads for a spherical region in $d=2$. (a) Two wedges with boundaries $A$ and $A^*$. (b) Gluing their red brane edges gives $M$. (c) The spatial CHM map sends the gluing curve to the throat $\E_n$. Purple threads cross once, orange threads return to the same boundary, and dashed gray threads approach the throat asymptotically. }
\label{fig:construction}
\end{figure*}

Next, we consider the canonical purification of the normalized power state $\rho_{A,n}$. Let $p_\alpha$ and $|\alpha\rangle_A$ be the nonzero eigenvalues and orthonormal eigenvectors of $\rho_A$, and let $|\alpha^*\rangle_{A^*}$ denote the corresponding vectors in the copy $A^*$. The canonical purification of $\rho_{A,n}$ is~\cite{Dutta:2019gen}
\begin{equation}
\begin{aligned}
 |\Psi_n\rangle
 &=\frac{1}{\sqrt{Z_n}}\sum_\alpha
 p_\alpha^{n/2}|\alpha\rangle_A|\alpha^*\rangle_{A^*},
\end{aligned}
 \label{eq:purification}
\end{equation}
with $\Tr_{A^*}|\Psi_n\rangle\langle\Psi_n|=\rho_{A,n}$. Thus $\wtS_n(A)=S(\rho_{A,n})$ is the entanglement entropy between $A$ and $A^*$. On the gravity side, the canonical purification is described by the reflected geometry obtained by gluing $W_n$ to its reflected copy $W_n^*$ along the cosmic brane $\mathcal{C}_{n}$ \cite{Dutta:2019gen,Gadde:2023classification},
\begin{equation}
 M=W_n\cup_{\mathcal{C}_{n}}W_n^*.
 \label{eq:reflected-double}
\end{equation}
For a single interval $A=(-R,R)$ in $d=2$, Fig.~\ref{fig:construction} illustrates the two wedges and their gluing in \eqref{eq:reflected-double}. We denote the spatial metric of $M$ by $h$, extending the metric on $W_n$ to $W_n^*$ by reflection. 

Applying the bulk CHM map to both \(W_n\) and \(W_n^*\), we obtain two copies of the exterior of the hyperbolic black hole \eqref{eq:spatial hyperbolic} glued along \(\E_n\), which we call the hyperbolic bridge. It is analogous to the Einstein--Rosen bridge of a two-sided eternal AdS black hole \cite{Maldacena:2001kr}. Let $(M_{\mathbb H},h_{\mathbb H})$ denote the CHM image of the reflected geometry $(M,h)$. Its throat $\E_n$ is the image of $\mathcal C_n$. Introduce a dimensionless proper distance $\ell$ from the throat by $|\ell|=\int_{r_n}^{\mathcal R_n(\ell)}dr/\sqrt{f_n(r)}$. The bridge metric is then
\begin{equation}
	M_{\mathbb H}=\mathbb R\times\mathbb H^{d-1},\quad
	h_{\mathbb H}=d\ell^2+\mathcal R_n(\ell)^2dH_{d-1}^2,
	\label{eq:bridge}
\end{equation}
where $\mathcal R_n(\ell)$ is smooth and even, with $\mathcal R_n(0)=r_n$. Both $\ell$ and $\mathcal R_n$ are dimensionless. Restoring units gives proper distance $L_{\rm AdS}\,d\ell$ and transverse radius $L_{\rm AdS}\mathcal R_n(\ell)$. The boundary ball radius $R$ enters the inverse CHM map and the boundary temperature, rather than the bulk metric prefactor. The two exteriors meet at $\ell=0$, while the asymptotic boundaries $\mathbb H_{L,R}^{d-1}$, reached as $\ell\to\pm\infty$, correspond to $A$ and $A^*$, respectively.

As an example, for $d=2$ we have $r_n=1/n$ and $\mathcal R_n(\ell)=\cosh\ell/n$. Since each transverse $\ell$-slice corresponds to $\mathbb H^1\simeq\mathbb R$ scaled by the factor \(\mathcal R_n(\ell)\), the transverse scale is smallest at $\ell=0$ and grows towards both asymptotic boundaries, as schematically shown in Fig.~\ref{fig:construction}(c).

\textcolor{blue}{\emph{Refined R\'enyi entropy from geodesic tensor network.}} We now construct the geodesic graph that perfectly tessellates the hyperbolic bridge $M_{\mathbb H}$, on which we define the tensor-network model that reproduces the holographic refined R\'enyi entropy. Following Refs.~\cite{Gu:2026rfo,Basu:2026hbg}, we shoot geodesics inward from the two boundaries using the intrinsic measure on each boundary surface. These geodesics interweave in the bulk and form the desired network. The intrinsic measure is the unique counting measure on intersections between geodesics and a hypersurface that is uniform over the surface and isotropic in all directions. Concretely, at an area element $dA$ on either boundary with unit normal $\v{n}$, choose the orientation of each geodesic intersecting the surface so that its unit tangent $\v{v}$ satisfies $\v{v}\cdot\v{n}>0$. The measure of intersections with $dA$ for directions in solid angle $d\Omega_{\v{v}}$ is then~\cite{Lin:2024fze,Basu:2026hbg,Santalo}
\begin{equation}
	d\Gamma=\frac{d-1}{\Omega_{d-2}}(\v{v}\cdot\v{n})\,dA \,d\Omega_{\v{v}},
	\quad  \v{v}\cdot\v{n}>0,
	\label{eq:local-pee}
\end{equation}
where $\v{v}\cdot\v{n}$ is the inner product in the spatial metric and $\Omega_{d-2}$ is the area of the unit sphere $S^{d-2}$.  This is the local intersection measure for boundary-sourced geodesics\footnote{A geodesic may intersect the source boundary more than once, so counting boundary intersections generally differs from counting distinct geodesics. See Refs.~\cite{Basu:2026hbg,Gu:2026rfo} for details.}. In the continuous geodesic network, the distribution of geodesics is described by vector fields, and counting their intersections with a surface is equivalent to computing the flux of these vector fields through it. Eq.~\eqref{eq:local-pee} is precisely such a flux formula. Since $\int_{\v{v}\cdot\v{n}>0}(\v{v}\cdot\v{n})\,d\Omega_{\v{v}}=\Omega_{d-2}/(d-1)$, integrating Eq.~\eqref{eq:local-pee} over inward directions gives $dA$, so the total intersection count reproduces the boundary area, in analogy with the Crofton formula.

We now examine geodesics on $M_{\mathbb H}$ with metric $h_{\mathbb H}$ in \eqref{eq:bridge}. Write $dH_{d-1}^2=\bar h_{ij}dq^i dq^j$, where $q^i$ are coordinates on unit hyperbolic space and $\bar h_{ij}$ is its metric. Parameterize a geodesic by arclength $\lambda$ in units of $L_{\rm AdS}$, and let a dot denote $d/d\lambda$. The projected geodesic equation is
\begin{equation}
 \frac{D\dot q^i}{d\lambda}
 +2\frac{\mathcal R_n'}{\mathcal R_n}\dot\ell\dot q^i=0,
 \label{eq:projected-geodesic}
\end{equation}
where $D/d\lambda$ is covariant differentiation for $\bar h_{ij}$ and $\mathcal R_n'=d\mathcal R_n/d\ell$. The second term is parallel to $\dot q^i$ and can be absorbed by reparameterizing the projected curve, so its image is a geodesic in $H^{d-1}$. Let $\chi$ denote arclength along the projected curve, defined by $d\chi^2=\bar h_{ij}\,dq^i dq^j$. The unit-speed condition gives
\begin{equation}
	\dot\ell^2+\frac{J^2}{\mathcal R_n^2}=1,\quad
	J=\mathcal R_n^2|\dot\chi|,
	\label{eq:first-integrals}
\end{equation}
where $J\geq0$ is conserved. The geodesics fall into three classes, as shown in Fig.~\ref{fig:construction}(c). For $0\leq J<r_n$, $\ell$ is monotonic and the geodesic connects the two boundaries, intersecting the throat exactly once. We call these crossing threads. For $J>r_n$, the geodesic turns at $\mathcal R_n=J$ and returns to the same boundary without reaching the throat. The critical geodesics with $J=r_n$ approach the throat asymptotically and form a set of zero measure.

Now we count the intersections between the geodesic graph and the throat \(\E_n\), the image of the cosmic brane. Since only the crossing threads intersect the throat, and each crosses \(\E_n\) exactly once with one endpoint on each boundary, the intersection number \(\mathcal N(\E_n)\) equals the number of crossing threads. It is therefore sufficient to count them from a single boundary. To do so, we evaluate the intrinsic measure \eqref{eq:local-pee} on a large constant-radius section \(r_c\) on the left boundary. Using \(\sin\theta=J/r_c\), the measure becomes \(d\Gamma\propto d\sigma_{\mathbb H}(q_L)J^{d-2}dJ\,d\Omega_{d-2}\), and the \(r_c\) dependence cancels as \(r_c\to\infty\). Carrying out the integrals over \(0\leq J<r_n\), the solid angle, and the left endpoint over the regulated hyperbolic volume \(V_{\mathbb H}\) then directly gives the intersection count at the throat
\begin{equation}
	\begin{aligned}
		\int_{\text{crossing}}d\Gamma=r_n^{d-1}V_{\mathbb H}=\Area(\E_n)=\mathcal N(\E_n),
	\end{aligned}
	\label{eq:throat-count}
\end{equation}
where \(\mathcal N(\E_n)\) is the number of crossing threads. 

We now consider the simplest PEE tensor-network model, the factorized one. Each geodesic represents an Einstein--Podolsky--Rosen (EPR) pair, with each qudit in the pair carrying a Hilbert space of dimension \(v\). The full geodesic network is therefore the tensor product of these EPR pairs. Since only the crossing threads connect \(A\) and \(A^*\), their total entanglement gives the entropy between the two copies in this network. With entanglement \(\log v=1/(4G_{\rm N})\) assigned to each EPR pair, the network entropy matches the holographic refined R\'enyi entropy,
\begin{equation}
	\wtS_n(A)=\frac{1}{4G_{\rm N}}\mathcal N(\E_n)
	=\frac{\Area(\E_n)}{4G_{\rm N}}
	=\frac{\Area(\mathcal{C}_{n})}{4G_{\rm N}}.
	\label{eq:main-result}
\end{equation}
Thus the factorized PEE tensor network reproduces the cosmic-brane prescription.

Equation~\eqref{eq:main-result} shows that the crossing threads reproduce the area of \(\mathcal{C}_n\), but it does not yet show that \(\E_n\) is the minimal cut of the geodesic network. To establish this, we compute the induced intersection measure on each constant-\(\ell\) hypersurface of the bridge, following Ref.~\cite{Gu:2026rfo}. The induced measure coincides with the intrinsic measure on every such hypersurface. Since every point of the bridge lies on one of them, the line density of the geodesics is uniform, and the same equality extends to an arbitrary hypersurface. Hence the region between \(\E_n\) and the boundary is perfectly tessellated by the geodesics emitted from that boundary. The intersection number of any surface is proportional to its area. Because \(\E_n\) is a minimal surface, its intersection number is minimal. We present the calculation in Appendix~\ref{app:minimal-cut}.

The canonical purification is not necessary, since the geodesics sourced from a single boundary are sufficient to reconstruct the area of $\mathcal{C}_n$. It is nevertheless useful because its bulk dual is the reflected double formed by gluing \(W_n\) to its reflection \(W_n^*\) along \(\mathcal C_n\). This casts \(\tilde S_n\) as an ordinary entanglement entropy between \(A\) and \(A^*\), makes the cosmic brane the gluing surface or throat, and shows that only crossing geodesics contribute.

\textcolor{blue}{\emph{Generic boundary regions.}} We now extend the construction to generic boundary regions. The boundary region \(A\) may be non-spherical or disconnected. We assume that the smooth \(n\)-sheeted replica manifold \(\mathcal B_n\) exists, with cosmic brane \(\mathcal C_n\) its fixed-point set. When several replica-symmetric saddles satisfy the same boundary conditions, the dominant saddle is selected by its Euclidean on-shell action, and we work with this saddle throughout. We also require that its time-reflection-symmetric wedge admits a smooth reflected double. At noninteger $n$, these assumptions apply to the continuation of the chosen saddle.

Let \(W_n\) be the region bounded by the replicated boundary and the cosmic brane \(\mathcal C_n\) on the chosen time-reflection-symmetric slice. Assume that the boundary-sourced geodesics, weighted by the intrinsic measure \eqref{eq:local-pee}, cover almost every unit-tangent direction in the reflected double and that there is no trapped set of positive geodesic measure. Then, according to Ref.~\cite{Gu:2026rfo}, these geodesics perfectly tessellate \(W_n\) in the measure-theoretic sense. A key technical ingredient is the transport theorem of Ref.~\cite{Gu:2026rfo} (see also Refs.~\cite{AlvarezPaiva:2008,AlvarezPaiva:2005,AlvarezPaiva:2006} for earlier discussion in the mathematical literature). This theorem states that the intrinsic intersection measure is preserved along the geodesic flow, so the local measure on a finite-radius section can be transported to the cosmic brane. Consequently, for every hypersurface \(\Sigma\subset W_n\), its intersection count equals its area, $\mathcal N(\Sigma)=\Area(\Sigma)$. To identify $\mathcal C_n$ with the minimal cut, we additionally require it to be globally area minimizing among the cuts homologous to $A$ in the reflected geometry. This is a separate geometric condition to be checked for the selected saddle. It does not follow from dominance of its on-shell action. Under this condition, the area identity makes \(\mathcal C_n\) the minimal cut of the geodesic network.

We now promote this geodesic network to a random PEE tensor network (see Appendix~\ref{app:pee-models} for details). We assign a random tensor to each vertex and contract along the geodesic links to obtain a boundary state. At large bond dimension $v$, with $\log v=1/(4G_{\rm N})$, the typical leading entanglement entropy of the replicated region $A$ is given by the minimal cut, provided the random-network concentration conditions hold~\cite{Hayden:2016cfa}. The minimal cut is the cosmic brane, so the resulting network entropy agrees at this order with \(S(\rho_{A,n})=\tilde S_n\),
\begin{equation}
	\tilde S_n=\frac{1}{4G_{\rm N}}\mathcal N(\mathcal C_n)
	=\frac{\Area(\mathcal C_n)}{4G_{\rm N}}.
	\label{eq:general-refined-count}
\end{equation}
Under the stated coverage and minimality assumptions, this reproduces the cosmic-brane prescription for generic boundary regions at leading order in the large-bond-dimension limit. The local counting measure and link states are independent of \(n\), while the changing geometry supplies the \(n\)-dependence of the matched refined R\'enyi entropy.

\textcolor{blue}{\emph{Discussion.}} In summary, we constructed an explicit family of geodesic tensor networks on replica geometries and reproduced the cosmic-brane prescription for refined R\'enyi entropy. The local counting measure and maximally entangled links are held fixed, while the graph changes with the replica saddle. This implements geometric backreaction in the network calculation and recovers the refined R\'enyi entropy's \(n\)-dependence. It gives a concrete graph realization of the fixed-area-state picture in Refs.~\cite{Dong:2018seb,Akers:2018fow}, in which different area saddles dominate at different \(n\). The construction thus identifies how the replica-dependent geometry controls the cut entropy without adjusting the link spectrum.

The construction above takes the bulk replica geometry as input, while a derivation of this geometry and its network from boundary data remains open. So far, the entropy matching has tested only the total cut entropy, leaving the finer entanglement structure of the network unexamined. To probe this structure, we compare the cross-copy boundary two-point PEE on $A\times A^*$ with the endpoint distribution of the bulk geodesic graph. For a single interval in CFT$_2$, we compute this distribution from the replica geodesic graph in Appendix~\ref{app:boundary-kernel} and reconstruct it from interval entropies in the canonical purification in Appendix~\ref{app:field-theory-pee}, within the additive linear combination (ALC) proposal~\cite{Wen:2018whg,Wen:2019iyq}. The inverse CHM map fixes the spatial ordering of the two copies. For dimensionless hyperbolic separation $\Delta$ between the endpoints, both calculations yield
\begin{equation*}
 \mathcal I_n^{\mathbb H}(\Delta)
 =\frac{c}{24n^2}\operatorname{sech}^2\!\left(\frac{\Delta}{2n}\right),
\end{equation*}
where $c$ is the CFT$_2$ central charge. Matching the full separation dependence tests how the network pairs degrees of freedom across the two copies, information that the total entropy alone does not fix. This agreement concerns the cross-copy sector. The same-copy sectors are not determined here. For higher-dimensional spherical regions, the geodesic graph also determines the cross-copy PEE structure, while extending the CFT reconstruction remains a question for future work.

\begin{acknowledgments}
The authors are supported by the NSFC Grant No.~12447108 and the Shing-Tung Yau Center of Southeast University.
\end{acknowledgments}

\bibliography{bib}

\clearpage

\section*{Supplemental Material}
\appendix
\setcounter{secnumdepth}{1}

\section{PEE tensor-network models}
\label{app:pee-models}

The geodesic graph that perfectly tessellates the bulk geometry is also called a partial-entanglement-entropy (PEE) network. It is the network of PEE threads. The PEE \(\mathcal I(X,Y)\) assigns an additive entanglement contribution to a pair of spacelike separated regions \(X\) and \(Y\) and satisfies the key property of additivity~\cite{Wen:2018whg,Wen:2019iyq,Han:2019scu} (see also Refs.~\cite{Chen:2014,Casini:2009} for related concepts). It admits a two-point density \(\mathcal I(x,y)\), so that \(\mathcal I(X,Y)=\int_X\int_Y \mathcal I(x,y)\). In the vacuum of a holographic CFT, the two-point PEE is geometrized by bulk geodesics called PEE threads~\cite{Lin:2023rxc,Lin:2024fze,Wen:2024uwr}, and \(\mathcal I(x,y)\) represents the density of geodesics connecting the boundary points \(x\) and \(y\). Each geodesic in the graph therefore represents a boundary two-point PEE, with a distribution determined by the PEE structure. This distribution coincides with the kinematic measure of integral geometry~\cite{Santalo,Helgason,Czech:2015qta} and hence yields the same geodesic graph that perfectly tessellates vacuum AdS space. In other words:
\begin{itemize}
\item \textit{The PEE threads are bulk geodesics labeled by points in kinematic space.}
\end{itemize}
Although the PEE structure motivates the construction, the main-text discussion does not rely on it. The essential ingredient is a geodesic graph that perfectly tessellates the relevant gravitational region.

We now briefly explain how to define tensor network models on such geodesic graphs (see Ref.~\cite{Wen:2025gui} for details). We assign a quantum state to each bulk site, which can be regarded as a vertex of the geodesic graph. At a site \(z\), we write
\begin{equation}
|T(z)\rangle=T_{u_1u_2\cdots}(z)|u_1\rangle_z|u_2\rangle_z\cdots,
\end{equation}
where each leg carries a \(v\)-dimensional qudit Hilbert space. The index $u_i=1,\ldots,v$ labels an orthonormal basis on the $i$th leg, and $T_{u_1u_2\cdots}(z)$ denotes the tensor coefficients in the product basis. The legs are continuously distributed and can be described by a vector field. In Fig.~\ref{fig:pee-models} only five legs are shown, while the blue region represents the remaining ones. Adjacent tensors are contracted along the PEE threads by projecting onto a maximally entangled state. The choice of vertex state \(|T(z)\rangle\) defines the model, and entanglement entropies can then be computed explicitly within the network.

\emph{(i) Factorized PEE network.} Each complete thread is a product of two-index tensors, each representing a pair of maximally entangled qudits (an EPR pair) of bond dimension \(v\)~\cite{Wen:2025gui}. The boundary state is therefore a tensor product of EPR pairs,
\begin{equation}
|\Psi\rangle=\bigotimes_{C}|\mathrm{EPR}\rangle_C,
\end{equation}
where \(C\) runs over all complete PEE threads. The entanglement entropy of a boundary region is the sum of the entropies of the EPR pairs whose threads connect the region to its complement. This model reproduces the RT formula for static single intervals and spherical regions, but not for generic disconnected or non-spherical regions, because the threads are decoupled and the required homologous cut may not exist.

\emph{(ii) HaPPY-like PEE network.} A perfect tensor is assigned to each site~\cite{Pastawski:2015qua}. A \(2k\)-index tensor is perfect if, for any bipartition of its legs into \(D\) and \(D^c\) with \(|D|\le |D^c|\), it is proportional to an isometry from \(D\) to \(D^c\). Here $k$ counts half the tensor legs and is unrelated to the replica index $n$. For the cut entropy to equal the number of cut bonds times $\log v$, the tensors on both sides of the cut must contract into isometries from the cut legs to the boundary legs. A greedy algorithm establishes this condition for suitable connected regions in HaPPY-type networks, but local perfectness alone does not guarantee it for an arbitrary geodesic graph. A quantum error-correcting interpretation also requires bulk input legs and a global encoding isometry. When these conditions hold, the Crofton identity identifies the minimal-cut entropy with the RT area.

\emph{(iii) Random PEE network.} A Haar-random state is assigned to each bulk site~\cite{Hayden:2016cfa}, and the entropy is computed after averaging over the random tensors. At large \(v\), the average is dominated by the minimal cut,
\begin{equation}
S(A)=\min_{\gamma_A}N(\gamma_A)\log v+\cdots,
\end{equation}
where $\gamma_A$ ranges over cuts homologous to $A$, $N(\gamma_A)$ is the cut-bond count, and the ellipsis denotes terms subleading at large $v$ for a fixed network. Using the Crofton formula \(N(\gamma_A)=\Area(\gamma_A)\) and identifying \(\log v=1/(4G_{\rm N})\), this reproduces the RT formula for generic boundary regions, including disconnected ones. Concentration bounds are additionally needed to infer that the averaged leading result holds for typical tensor choices. The approximation requires the finite-$v$ corrections to be controlled~\cite{Hayden:2016cfa}.

The three models are summarized in Table~\ref{tab:pee-models}. The factorized model is the simplest and most transparent; the HaPPY-like model adds an isometric encoding and quantum error correction when the stated contraction conditions hold; the random model gives the leading RT formula for more general regions at large $v$, subject to concentration.

\par\medskip\noindent\begin{minipage}{\columnwidth}
\centering
\small
\setlength{\tabcolsep}{4pt}
\begin{tabular}{@{}l l l@{}}
\hline\hline
Model & Vertex state & Regime \\
\hline
Factorized & EPR pairs & single intervals and spheres \\
HaPPY-like & perfect tensors & isometric cuts \\
Random & Haar-random & generic regions for large \(v\) \\
\hline\hline
\end{tabular}
\captionof{table}{Three PEE tensor-network models.}
\label{tab:pee-models}
\end{minipage}\par\medskip

\par\medskip\noindent\begin{minipage}{\columnwidth}
	\centering
	{\begin{tikzpicture}[
			x=1cm,y=1cm,
			line cap=round,line join=round,
			font=\small,
			pee thread/.style={draw=PEEThread,line width=.85pt},
			pee end/.style={circle,draw=PEEInk,fill=white,line width=.65pt,
				minimum size=3.8mm,inner sep=0pt}
			]
			\definecolor{PEEThread}{RGB}{12,116,173}
			\definecolor{PEECloud}{RGB}{216,234,247}
			\definecolor{PEEInk}{RGB}{38,43,48}
			
			\coordinate (pee z) at (-.48,-.05);
			\coordinate (pee zp) at (2.00,1.12);
			\coordinate (pee a) at (-2.45,.72);
			\coordinate (pee b) at (-.55,1.65);
			\coordinate (pee d) at (.60,-1.60);
			\coordinate (pee e) at (-1.90,-1.20);
			
			\fill[PEECloud] plot[smooth cycle,tension=.72] coordinates {
				(-2.45,.72) (-1.60,1.45) (-.55,1.65)
				(1.10,1.54) (2.00,1.12) (2.60,-.38)
				(1.85,-1.24) (.60,-1.60) (-1.00,-1.52)
				(-1.90,-1.20) (-2.50,-.25)
			};
			
			\draw[pee thread]
			(pee z) .. controls (-1.093333,.55) and (-1.75,.806667) .. (pee a);
			\draw[pee thread]
			(pee z) .. controls (-.293333,.53) and (-.316667,1.096667) .. (pee b);
			\draw[pee thread]
			(pee z) .. controls (.173333,-.483333) and (.533333,-1.00) .. (pee d);
			\draw[pee thread]
			(pee z) .. controls (-.726667,-.683333) and (-1.20,-1.066667) .. (pee e);
			\draw[pee thread]
			(pee z) .. controls (.473333,.15) and (1.30,.54) .. (pee zp)
			coordinate[pos=.55] (pee pair left)
			coordinate[pos=.60] (pee pair right);
			
			\foreach \site in {a,b,d,e}{
				\node[pee end] at (pee \site) {};
			}
			\node[circle,draw=PEEInk,fill=white,line width=.75pt,
			minimum size=6.5mm,inner sep=0pt] at (pee z) {$z$};
			\node[circle,draw=PEEInk,fill=white,line width=.70pt,
			minimum size=5.6mm,inner sep=0pt] at (pee zp) {$z'$};
			\fill[PEEInk] (pee pair left) circle[radius=1.15pt];
			\fill[PEEInk] (pee pair right) circle[radius=1.15pt];
			\node[inner sep=1pt] at (.77,.96) {$|z z'\rangle$};
	\end{tikzpicture}}

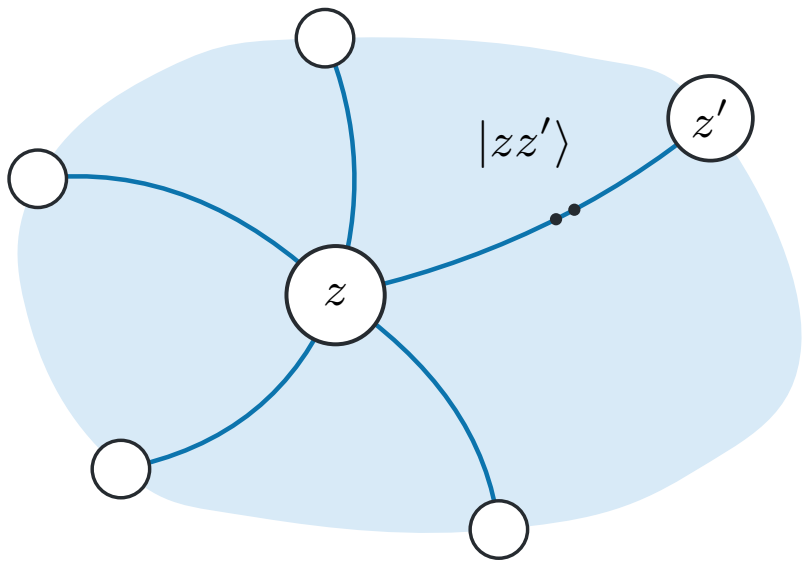
\captionof{figure}{A tensor state located at a bulk site \(z\). The legs are continuously distributed and only five legs are explicitly shown. The contraction between tensors at \(z\) and an adjacent site \(z'\) is realized by projecting onto a maximally entangled state $\ket{z z'}$ along the geodesic connecting them.}
\label{fig:pee-models}
\end{minipage}\par\medskip

\emph{Fixed-area flatness and replica-dependent networks.} At a finite regulator, let $\sigma_{A,1}$ be the reduced state of region $A$ in the factorized PEE network at $n=1$. Tracing out its complement leaves a maximally mixed state from the EPR pairs crossing the cut, tensored with the pure state of the EPR pairs entirely inside $A$,
\begin{equation}
 \sigma_{A,1}=\frac{\mathbf 1_{D_1}}{D_1}\otimes|\chi_1\rangle\langle\chi_1|
 =\frac{P_1}{D_1}.
 \label{eq:factorized-reduced-state}
\end{equation}
Here $D_1=v^{N_1}$, where $N_1$ counts the cut EPR pairs and $v$ is their bond dimension. The state $|\chi_1\rangle$ is the normalized product of the internal EPR pairs, and $P_1=\mathbf 1_{D_1}\otimes|\chi_1\rangle\langle\chi_1|$ is the support projector, with $\Tr P_1=D_1$. Consequently,
\begin{equation}
 \frac{\sigma_{A,1}^{\,n}}{\Tr\sigma_{A,1}^{\,n}}
 =\sigma_{A,1},\qquad n>0.
 \label{eq:factorized-power-invariance}
\end{equation}
Thus the factorized model gives an exact network realization of the flat-spectrum mechanism associated with fixed-area states~\cite{Akers:2018fow,Dong:2018seb}.

Our construction instead allows the geometry and its geodesic network to vary with $n$, following the gravitational replica saddle. Let $\sigma_{A,n}$ denote the reduced state produced by the network on the $n$-dependent geometry. With the entropy normalization specified above, its cut entropy reproduces the cosmic-brane area, $S(\sigma_{A,n})\simeq\Area(\mathcal C_n)/(4G_{\rm N})=\wtS_n(A)$, at leading semiclassical order and within the regime of validity of the chosen model. The changing geometry therefore recovers the $n$-dependence of the refined R\'enyi entropy even though the individual EPR links remain maximally entangled. Here $n$ labels distinct networks, and this entropy matching does not require $\sigma_{A,n}=\sigma_{A,1}^{\,n}/\Tr\sigma_{A,1}^{\,n}$.

\section{CHM map and canonical purification}
\label{app:chm-purification}

Throughout this Appendix we set \(R=1\), so that \(\Lambda(x)=(1-|x|^2)/2\). We also retain the main-text convention $L_{\rm AdS}=1$. The Hamiltonian $\H$ below generates dimensionless hyperbolic time. For general $R$, the physical Hamiltonian is $\H/R$ and the physical inverse temperature is $2\pi nR$.

Let $\H$ generate dimensionless hyperbolic time and $U_A$ implement the boundary CHM transformation~\cite{Casini:2011kv}. Then
\begin{equation}
 U_A\rho_AU_A^\dagger=\frac{e^{-2\pi\H}}{\mathcal Z(2\pi)},
 \quad \mathcal Z(b)=\Tr e^{-b\H}.
 \label{eq:chm-state}
\end{equation}
Here $b$ is the dimensionless inverse temperature. At $R=1$ it equals the physical inverse temperature in the chosen units. Taking the $n$-th power multiplies the exponent by $n$. Taking the trace and dividing by $Z_n$ gives
\begin{equation}
 \begin{aligned}
 Z_n&=\frac{\mathcal Z(2\pi n)}{\mathcal Z(2\pi)^n},\\
 U_A\rho_{A,n}U_A^\dagger&=\frac{e^{-2\pi n\H}}{\mathcal Z(2\pi n)}.
 \end{aligned}
 \label{eq:chm-powered-state}
\end{equation}
Thus the normalized power state becomes a thermal state of the same Hamiltonian at inverse temperature $\beta_n=2\pi n$. In the hyperbolic frame, let $L,R$ label the system and its conjugate auxiliary copy. The CHM map sends the eigenvectors of $\rho_A$ to energy eigenstates, $|E_\alpha\rangle_L=U_A|\alpha\rangle_A$, with $p_\alpha=e^{-2\pi E_\alpha}/\mathcal Z(2\pi)$. Applying the conjugate map $U_A^*$ to the auxiliary factor in \eqref{eq:purification} gives
\begin{equation}
 \begin{aligned}
 |\Psi_n\rangle_{LR}&=(U_A\otimes U_A^*)|\Psi_n\rangle\\
 &=\frac{\sum_\alpha e^{-\pi n E_\alpha}|E_\alpha\rangle_L
 |E_\alpha^*\rangle_R}{\sqrt{\mathcal Z(2\pi n)}}.
 \end{aligned}
 \label{eq:tfd}
\end{equation}
This is a thermofield double, so applying CHM to both copies is compatible with taking the canonical purification. Its Euclidean preparation over an interval of length $\beta_n/2$ gives the two-sided black-hole interpretation~\cite{Maldacena:2001kr}. The right trace agrees with \eqref{eq:chm-powered-state}.

Let $\Phi$ denote a bulk extension of the spatial CHM map. Choose bulk coordinates $(\tilde z,x)$, where $\tilde z>0$ vanishes at the conformal boundary. The inherited spatial metric on $W_n$ takes the form~\cite{Agon:2025fluids}
\begin{equation}
 h\big|_{W_n}=\frac{1}{\tilde z^2}\left[
 dx\cdot dx+
 \frac{\bigl(d\tilde z-\tilde z\,d\ln\Lambda(x)\bigr)^2}
 {1-\tilde z^2/\Lambda(x)^2-\mu_n\tilde z^d/\Lambda(x)^d}
 \right],
 \label{eq:finite-n-wedge-chart}
\end{equation}
with $|x|<R$ and $0<\tilde z\leq\Lambda(x)/r_n$. The map $\Phi|_{W_n}:(\tilde z,x)\mapsto(r,q)$, where $q=(\eta,\hat x)$, is
\begin{equation}
 r=\frac{\Lambda(x)}{\tilde z},\quad
 \eta=2\operatorname{arctanh}\frac{|x|}{R},\quad
 \hat x=\frac{x}{|x|},
 \label{eq:finite-n-bulk-chm-map}
\end{equation}
with the angular coordinate unnecessary at $x=0$. The identities $dr/r=d\ln\Lambda(x)-d\tilde z/\tilde z$ and $r^2dH_{d-1}^2=dx\cdot dx/\tilde z^2$ transform \eqref{eq:finite-n-wedge-chart} into \eqref{eq:spatial hyperbolic}. Since $\C_n$ is located at $\tilde z=\Lambda(x)/r_n$, the map gives $\Phi(\C_n)=\E_n$. To see that the isometry extends regularly to this surface, use the Euclidean metric $ds_E^2=f_n(r)d\tau^2+h_{\mathbb H}$ and introduce $\varrho=\sqrt{2n(r-r_n)}$. Near the horizon,
\begin{equation}
 \begin{aligned}
 f_n(r)&=\frac{2}{n}(r-r_n)+O((r-r_n)^2),\\
 ds_E^2&\simeq d\varrho^2+\frac{\varrho^2}{n^2}d\tau^2+r_n^2dH_{d-1}^2.
 \end{aligned}
 \label{eq:brane-tip-metric}
\end{equation}
The angular coordinate $\tau/n$ has period $2\pi$ in the smooth black hole and $2\pi/n$ in the replica quotient, where the cosmic brane sits at the tip~\cite{Dong:2016fnf}. The fixed-$\tau$ spatial metric is regular at $\varrho=0$, and the induced metric is $ds_{\E_n}^2=r_n^2dH_{d-1}^2$. With the cutoff on $\C_n$ chosen as the inverse image of the cutoff on $\E_n$, the isometry gives \eqref{eq:brane-regulated-area}. The different Euclidean time periods affect only the transverse opening angle and introduce no factor of $n$ in this spatial area equality.

\section{Minimality of the throat cut}
\label{app:minimal-cut}

In this Appendix we set \(R=1\) and retain \(L_{\rm AdS}=1\). We count each complete unoriented geodesic once, identifying the two endpoint representations of returning geodesics~\cite{Gu:2026rfo}. We show that the boundary-sourced geodesics induce the intrinsic measure on every constant-\(\ell\) hypersurface of the bridge, and hence on any hypersurface in the region between the boundary and the throat.

Consider the dimensionless bridge metric
\begin{equation}
h_{\mathbb H}=d\ell^2+\mathcal R_n(\ell)^2dH_{d-1}^2,
\end{equation}
with \(\mathcal R_n(0)=r_n\) and \(\mathcal R_n\) even. At a constant-\(\ell\) slice \(\Sigma_\ell\simeq H^{d-1}\), let \(\alpha\) be the angle between a geodesic and the normal \(\partial_\ell\). The conserved momentum satisfies
\begin{equation}
\sin\alpha=\frac{J}{\mathcal R_n(\ell)}.
\end{equation}
Adapting Sec.~2.2 of Ref.~\cite{Gu:2026rfo}, we first use the isometries of the bridge. For fixed \(\ell\), the slice \(\Sigma_\ell\simeq H^{d-1}\) is homogeneous and isotropic. The isometry group of \(H^{d-1}\) acts transitively on it, and the boundary source is invariant under these isometries. Consequently, for fixed \(J\) and a chosen radial branch, the induced measure on \(\Sigma_\ell\) must be proportional to the unique invariant measure on the slice, \(d\sigma_{\mathbb H}(q)\,d\Omega_{d-2}\). Here $d\sigma_{\mathbb H}(q)$ is the volume element of the unit hyperbolic metric, $q$ is a point on that space, and $d\Omega_{d-2}$ is the solid-angle element of the projected direction. Write $c_d=(d-1)/\Omega_{d-2}$ for the normalization in \eqref{eq:local-pee}. Writing the fixed-\(J\) measure as \(C_\ell(J)\,dJ\,d\sigma_{\mathbb H}(q)d\Omega_{d-2}\), the coefficient \(C_\ell(J)\) is fixed by the boundary normalization \eqref{eq:local-pee}. As in Ref.~\cite{Gu:2026rfo}, the total fixed-\(J\) measure is independent of \(\ell\), so matching it to the boundary gives
\begin{equation}
C_\ell(J)=\frac{d-1}{\Omega_{d-2}}J^{d-2}.
\end{equation}
Thus
\begin{equation}
d\Gamma^{\rm ind}_{\Sigma_\ell}
=\frac{d-1}{\Omega_{d-2}}
d\sigma_{\mathbb H}(q)\,J^{d-2}dJ\,d\Omega_{d-2}.
\end{equation}
Using \(J=\mathcal R_n(\ell)\sin\alpha\) and \(dA_\ell=\mathcal R_n(\ell)^{d-1}d\sigma_{\mathbb H}(q)\), this becomes
\begin{equation}
d\Gamma^{\rm ind}_{\Sigma_\ell}
=\frac{d-1}{\Omega_{d-2}}
\cos\alpha\,dA_\ell\,\sin^{d-2}\alpha\,d\alpha\,d\Omega_{d-2}.
\end{equation}
Since \(d\Omega_{\bf v}=\sin^{d-2}\alpha\,d\alpha\,d\Omega_{d-2}\) and \(v\cdot n=\cos\alpha\), this is exactly the intrinsic measure \eqref{eq:local-pee} on \(\Sigma_\ell\). Thus the induced and intrinsic measures agree on every constant-\(\ell\) slice.

The two boundary sources together supply almost every geodesic direction throughout the bridge. One source suffices between that boundary and the throat, while returning directions in the other exterior are supplied by the other source. To extend this result to an arbitrary hypersurface, define the line density \(\rho(P,[v])\) as the density of geodesic intersections with a unit area transverse to \(v\) at \(P\), per unit solid angle in the direction \([v]\). The class $[v]$ identifies the opposite unit tangents $v$ and $-v$. Choose its local representative with $v\cdot n>0$. Equivalently, for an area element \(dA_\Sigma\) with unit normal \(n\), the induced measure is
\begin{equation}
d\Gamma^{\rm ind}_{\Sigma}
=\rho(P,[v])\,(v\cdot n)\,dA_\Sigma\,d\Omega_{\bf v}.
\end{equation}
Since every point of the bridge lies on some \(\Sigma_\ell\), and on each \(\Sigma_\ell\) the induced measure equals the intrinsic measure, we have \(\rho(P,[v])=c_d=(d-1)/\Omega_{d-2}\) throughout the region. Therefore, for any hypersurface \(\Sigma\),
\begin{equation}
d\Gamma^{\rm ind}_{\Sigma}
=c_d(v\cdot n)\,dA_\Sigma\,d\Omega_{\bf v}
=d\Gamma^{\rm intr}_{\Sigma},
\end{equation}
and hence \(\mathcal N(\Sigma)=\Area(\Sigma)\). The entire region between the boundary and the throat is therefore perfectly tessellated.

Finally, the area of \(\Sigma_\ell\) is \(\mathcal R_n(\ell)^{d-1}V_{\mathbb H}\), which is minimized at \(\ell=0\) because \(\mathcal R_n\) is even and increases away from the throat. Thus \(\E_n=\Sigma_0\) is a minimal surface. Since its intersection number equals its area and all homologous surfaces have larger or equal intersection number, \(\E_n\) is the minimal cut.

\section{Two-point PEE from the replica geodesic graph}
\label{app:boundary-kernel}

We extract the cross-copy two-point PEE predicted by the replica geodesic graph for a vacuum spherical region $A=\{x\in\mathbb R^{d-1}:|x|<R\}$ in CFT$_d$. The state is the canonical purification $|\Psi_n\rangle$ of $\rho_{A,n}$ in \eqref{eq:purification}, with $n>0$ on the Einstein-gravity branch \eqref{eq:spatial hyperbolic}. We first determine the density of geodesics connecting $A$ to its auxiliary copy $A^*$, then assign their entanglement weight using the PEE-thread prescription~\cite{Lin:2023rxc,Lin:2024fze}. This determines the $A\times A^*$ sector of the boundary PEE structure. Same-copy pairs in $A\times A$ or $A^*\times A^*$ are not reconstructed here. Appendix~\ref{app:field-theory-pee} independently reconstructs the single-interval result from boundary entropies within the ALC proposal.

Under CHM, $A$ and $A^*$ map to the two boundaries $\mathbb H_L^{d-1}$ and $\mathbb H_R^{d-1}$ of the bridge \eqref{eq:bridge}. In this Appendix we retain the boundary ball radius $R$ and use $L_{\rm AdS}=1$ as in the main text. The variables $r$, $r_n$, $\ell$, and $\mathcal R_n(\ell)$, as well as the metric $dH_{d-1}^2$ on unit hyperbolic space, are dimensionless. Restoring the bulk curvature radius multiplies the bridge metric by $L_{\rm AdS}^2$ and physical areas by $L_{\rm AdS}^{d-1}$. These factors are independent of $R$. Here $\ell$ is proper distance from the throat in AdS units, $\mathcal R_n(\ell)$ is the radius profile in \eqref{eq:bridge}, and $r_n=\mathcal R_n(0)$ is the dimensionless throat radius. The exterior radial coordinate $r$ and blackening function $f_n(r)$ are defined in \eqref{eq:spatial hyperbolic}. Write $d\sigma_{\mathbb H}(q)$ for the volume element of $dH_{d-1}^2$.

Reflection identifies the coordinate frames of the two boundaries. For endpoints $q_L\in\mathbb H_L^{d-1}$ and $q_R\in\mathbb H_R^{d-1}$, $\Delta(q_L,q_R)$ denotes the unit hyperbolic distance between their coordinate values in this common frame. Thus $\Delta=0$ labels matching coordinates on distinct boundaries.

Only the crossing geodesics, with one endpoint on each boundary, contribute to the PEE between the copies. Their conserved momentum is $J=\mathcal R_n^2|d\chi/d\lambda|$, as in \eqref{eq:first-integrals}, where $\lambda$ is arclength in the dimensionless bridge metric and $\chi$ is arclength along the projected hyperbolic geodesic. For $0\leq J<r_n$, their endpoint separation is
\begin{equation}
 \Delta(J)=2\int_{r_n}^{\infty}
 \frac{J\,dr}{r^2\sqrt{f_n(r)}\sqrt{1-J^2/r^2}}.
 \label{eq:separation}
\end{equation}
Indeed, \eqref{eq:first-integrals} and $dr/d\ell=\pm\sqrt{f_n(r)}$ imply, on the two exterior branches,
\begin{equation}
 \frac{d\chi}{dr}
 =\pm\frac{J}{r^2\sqrt{f_n(r)}\sqrt{1-J^2/r^2}}.
 \label{eq:radial-separation-derivative}
\end{equation}
Each branch contributes half of \eqref{eq:separation}. Differentiating its integrand gives $\Delta'(J)>0$, with $\Delta(0)=0$. The near-throat expansion \eqref{eq:brane-tip-metric} gives a logarithmic divergence as $J\to r_n^-$. The inverse function $J(\Delta)$ is therefore uniquely defined for every finite $\Delta\geq0$.

To obtain the endpoint measure, evaluate the intrinsic intersection measure \eqref{eq:local-pee} on a constant-radius section $r=r_c$ in the left exterior. Its area element in AdS units is $r_c^{d-1}d\sigma_{\mathbb H}$. Let $\theta$ be the angle between the inward unit tangent $\mathbf v$ and the inward normal. The momentum relation gives $\sin\theta=J/r_c$, while the directional solid angle decomposes as $d\Omega_{\mathbf v}=\sin^{d-2}\theta\,d\theta\,d\Omega_{d-2}$. Here $d\Omega_{d-2}$ is the angular measure on the unit sphere $S^{d-2}$ of projected directions, with total area $\Omega_{d-2}$. Since $r_c^{d-1}\cos\theta\sin^{d-2}\theta\,d\theta=J^{d-2}dJ$, the boundary limit gives
\begin{equation}
 d\Gamma=\frac{(d-1)}{\Omega_{d-2}}
 d\sigma_{\mathbb H}(q_L)\,J^{d-2}dJ\,d\Omega_{d-2}.
 \label{eq:boundary-momentum-measure}
\end{equation}
The section coordinate approaches $q_L$ as $r_c\to\infty$, and all powers of $r_c$ cancel. The measure $d\Gamma$ counts unoriented crossing geodesics, each represented once by its inward orientation at the left boundary.

For fixed $q_L$, introduce geodesic polar coordinates on $\mathbb H_R^{d-1}$ about its reflected image. Then $d\sigma_{\mathbb H}(q_R)=\sinh^{d-2}\!\Delta\,d\Delta\,d\Omega_{d-2}$. Define the geometric endpoint density $\mathcal J_n^{\mathbb H}$ by $d\Gamma=\mathcal J_n^{\mathbb H}(\Delta)d\sigma_{\mathbb H}(q_L)d\sigma_{\mathbb H}(q_R)$. Changing variables from $J$ to $\Delta$ yields
\begin{equation}
 \mathcal J_n^{\mathbb H}(\Delta)
 =\frac{(d-1)}{\Omega_{d-2}}
 \left(\frac{J(\Delta)}{\sinh\Delta}\right)^{d-2}
 \frac{dJ(\Delta)}{d\Delta}.
 \label{eq:geodesic-endpoint-density}
\end{equation}
At $\Delta=0$, this expression is understood by continuity. In the PEE network, each thread carries entropy $1/(4G_{\rm N})$, where $G_{\rm N}$ is the bulk Newton constant. The two-point PEE density with respect to the two unit hyperbolic volume elements is therefore
\begin{equation}\label{eq:explicit-kernel}
 \mathcal I_n^{\mathbb H}(\Delta)
 =\frac{\mathcal J_n^{\mathbb H}(\Delta)}{4G_{\rm N}}
 =\frac{(d-1)}{4G_{\rm N}\Omega_{d-2}}
 \left(\frac{J(\Delta)}{\sinh\Delta}\right)^{d-2}
 \frac{dJ(\Delta)}{d\Delta}.
\end{equation}
Thus $\mathcal I_n^{\mathbb H}$ is the entropy-weighted endpoint density, whereas $\mathcal J_n^{\mathbb H}$ is the geometric counting density. There is no additional factor $1/2$, since fixing the left endpoint already counts each crossing thread once. Integrating over the full right boundary gives
\begin{equation}
 \begin{aligned}
 \int_{\mathbb H_R^{d-1}}d\sigma_{\mathbb H}(q_R)\,\mathcal I_n^{\mathbb H}(\Delta)
 &=\frac{(d-1)}{4G_{\rm N}}\int_0^{r_n}J^{d-2}\,dJ\\
 &=\frac{r_n^{d-1}}{4G_{\rm N}}.
 \end{aligned}
 \label{eq:kernel-normalization}
\end{equation}
Regulate the remaining left integral by the unit hyperbolic volume $V_{\mathbb H}$ defined below \eqref{eq:brane-regulated-area}. The total PEE is then $r_n^{d-1}V_{\mathbb H}/(4G_{\rm N})=\wtS_n(A)$, in agreement with \eqref{eq:main-result}. This prescription regulates one endpoint integral after taking the full marginal over the other copy. It counts every crossing thread incident on the regulated left region, wherever its other endpoint lies. Restricting both endpoint integrals instead retains only threads with both endpoints in the regulated regions and therefore defines a different finite-cutoff quantity. We use the former prescription consistently in comparing with the horizon area.

To express the density on $A\cup A^*$, use the same inverse spatial CHM map \eqref{eq:chm-spatial-map} on both copies. The coordinates $x\in A$ and $y\in A^*$ are independent vectors in the ball $|x|,|y|<R$. With $\Lambda(s)=(R^2-|s|^2)/(2R)$ for $s=x,y$, the endpoint volume elements obey $d\sigma_{\mathbb H}(q_L)=\Lambda(x)^{-(d-1)}d^{d-1}x$ and similarly for $q_R,y$. Invariance of the PEE endpoint measure gives
\begin{equation}
 \mathcal I_n(x,y)=\frac{\mathcal I_n^{\mathbb H}\!\bigl(\Delta(x,y)\bigr)}
 {\Lambda(x)^{d-1}\Lambda(y)^{d-1}}.
 \label{eq:pre-copy-coordinates}
\end{equation}
Here $\mathcal I_n(x,y)$ is the density with respect to $d^{d-1}x\,d^{d-1}y$, and $\Delta(x,y)=\Delta(q(x),q(y))$. To evaluate it, embed unit hyperbolic space as $X(q)=(\cosh\eta,\sinh\eta\,\hat x)$, using the radial and angular coordinates $\eta,\hat x$ of \eqref{eq:chm-spatial-map}. The identity $\cosh\Delta=-X(q_L)\cdot_{\rm M}X(q_R)$, with Minkowski signature $(-,+,\ldots,+)$, gives
\begin{equation}
 \begin{aligned}
 \cosh\Delta(x,y)&=1+\frac{|x-y|^2}{2\Lambda(x)\Lambda(y)},\\
 \Delta(x,y)&=2\operatorname{arsinh}
 \frac{|x-y|}{2\sqrt{\Lambda(x)\Lambda(y)}}.
 \end{aligned}
 \label{eq:chm-distance}
\end{equation}
The norm $|x-y|$ and the dot product below refer to the common Euclidean coordinate frame of the two balls.

\emph{Vacuum limit and a single interval.}

At $n=1$, the black hole is the hyperbolic slicing of pure AdS, with $r_1=1$, $\mu_1=0$, and $f_1(r)=r^2-1$. Here $\mu_n$ is the mass parameter in \eqref{eq:spatial hyperbolic}. In \eqref{eq:separation}, the substitution $u=\sqrt{1-r^{-2}}$ gives $\Delta(J)=2\operatorname{arctanh}J$. Hence
\begin{equation}
 \begin{aligned}
 J(\Delta)&=\tanh\frac\Delta2,\\
 \mathcal I_1^{\mathbb H}(\Delta)
 &=\frac{(d-1)}{4G_{\rm N}\Omega_{d-2}\,2^{d-1}\cosh^{2(d-1)}(\Delta/2)}.
 \end{aligned}
 \label{eq:vacuum-separation}
\end{equation}
Substituting this result and \eqref{eq:chm-distance} into \eqref{eq:pre-copy-coordinates} yields
\begin{equation}
 \mathcal I_1(x,y)=\frac{(d-1)\,2^{d-1}R^{2(d-1)}}
 {4G_{\rm N}\Omega_{d-2}
 \bigl(R^4-2R^2x\cdot y+|x|^2|y|^2\bigr)^{d-1}}.
 \label{eq:vacuum-copy-kernel}
\end{equation}

To compare with the vacuum PEE between $A$ and its exterior $\bar A=\{y':|y'|>R\}$, map the auxiliary point to the exterior by spherical inversion,
\begin{equation}
 y'=\frac{R^2y}{|y|^2},\quad |y'|>R,\quad
 d^{d-1}y'=\left(\frac{R^2}{|y|^2}\right)^{d-1}d^{d-1}y.
 \label{eq:auxiliary-inversion}
\end{equation}
The point $y=0$ maps to infinity. Using $|x-y'|^2=(R^4-2R^2x\cdot y+|x|^2|y|^2)/|y|^2$ gives the measure equality
\begin{equation}
 \begin{aligned}
 \mathcal I_1(x,y)\,d^{d-1}y
 &=\mathcal I_{\rm vac}(x,y')\,d^{d-1}y',\\
 \mathcal I_{\rm vac}(x,y')
 &=\frac{(d-1)\,2^{d-1}}{4G_{\rm N}\Omega_{d-2}|x-y'|^{2(d-1)}}.
 \end{aligned}
 \label{eq:vacuum-recovery}
\end{equation}
Thus the graph prescription recovers the standard holographic vacuum two-point PEE $\mathcal I_{\rm vac}$~\cite{Lin:2023rxc,Lin:2024fze}.

For $d=2$, consider the interval $A=(-R,R)$ at general $n>0$. The bridge has $r_n=1/n$ and $\mathcal R_n(\ell)=\cosh\ell/n$, as in the main text~\cite{Banados:1992wn}. Evaluating \eqref{eq:separation} and using \eqref{eq:explicit-kernel}, $\Omega_0=2$, and the Brown--Henneaux relation $c=3L_{\rm AdS}/(2G_{\rm N})=3/(2G_{\rm N})$~\cite{Brown:1986nw}, where $c$ is the CFT$_2$ central charge, gives
\begin{equation}
 \begin{aligned}
 J(\Delta)&=\frac1n\tanh\frac{\Delta}{2n},\\
 \mathcal I_n^{\mathbb H}(\Delta)
 &=\frac{c}{24n^2}\operatorname{sech}^2\!\left(\frac{\Delta}{2n}\right).
 \end{aligned}
 \label{eq:btz-kernel}
\end{equation}
This agrees with the BTZ thread density~\cite{Basu:2026vsq,Lin:2025jjh}. In one spatial dimension, use the signed CHM coordinate $q(s)=\log[(R+s)/(R-s)]$ for $s=x,y$, so $q_L=q(x)$, $q_R=q(y)$, and $\Delta=|q(x)-q(y)|$. The same coordinate function on both copies follows from reflection. Their opposite orientations in the doubled system are explained in Appendix~\ref{app:field-theory-pee}. Since $dq/ds=\Lambda(s)^{-1}=2R/(R^2-s^2)$, \eqref{eq:pre-copy-coordinates} becomes
\begin{equation}
 \mathcal I_n(x,y)
 =\frac{cR^2}{6n^2(R^2-x^2)(R^2-y^2)}
 \operatorname{sech}^2\!\left(\frac{q(x)-q(y)}{2n}\right).
 \label{eq:interval-pre}
\end{equation}
Equations~\eqref{eq:btz-kernel} and \eqref{eq:interval-pre} are precisely the densities reconstructed from boundary interval entropies in Appendix~\ref{app:field-theory-pee}.

Integrating over the full auxiliary interval $A^*=(-R,R)$, or equivalently $q_R\in\mathbb R$, gives the marginal density on $A$,
\begin{equation}
 \int_{A^*}dy\,\mathcal I_n(x,y)
 =\frac{cR}{3n(R^2-x^2)}.
 \label{eq:interval-copy-marginal}
\end{equation}
Implement the one-endpoint regulator used above by restricting $x$ to $(-R+\epsilon,R-\epsilon)$, where $\epsilon>0$ is the boundary ultraviolet cutoff, while keeping the full $y$ integral. Then
\begin{equation}
 \begin{aligned}
 \int_{-R+\epsilon}^{R-\epsilon}dx\int_{A^*}dy\,\mathcal I_n(x,y)
 &=\frac{c}{3n}\log\frac{2R-\epsilon}{\epsilon}\\
 &=\frac{c}{3n}\log\frac{2R}{\epsilon}+o(1).
 \end{aligned}
 \label{eq:interval-copy-integral}
\end{equation}
Here $o(1)$ vanishes as $\epsilon/R\to0$. The auxiliary integral is kept complete before the cutoff is imposed on $x$, as required by the preceding one-endpoint prescription. Retaining the universal logarithm gives $\wtS_n(A)$. Integrating the refined R\'enyi relation \eqref{eq:refined-power}, with $S_n(A)$ finite at $n=1$, gives the ordinary R\'enyi entropy,
\begin{equation}
 \wtS_n(A)=\frac{c}{3n}\log\frac{2R}{\epsilon},\quad
 S_n(A)=\frac{c}{6}\left(1+\frac1n\right)\log\frac{2R}{\epsilon}.
 \label{eq:interval-entropies}
\end{equation}
These universal terms agree with the single-interval CFT$_2$ result~\cite{Calabrese:2004eu}.

\emph{Large-separation decay.} The higher-dimensional kernel also determines how replica backreaction changes the endpoint pairing. At fixed finite $n>0$, the regularity condition $f_n'(r_n)=2/n$ gives $\mathcal R_n(\ell)=r_n+\ell^2/(2n)+O(\ell^4)$ near the throat. Crossing geodesics with $J\to r_n^-$ spend an increasingly long distance near this minimum. Writing $\kappa_n=\sqrt{r_n/n}$, their separation has the asymptotic form
\begin{equation}
 \Delta(J)=-\frac{1}{\kappa_n}\log\!\left(1-\frac{J}{r_n}\right)+O(1),
 \qquad J\to r_n^-.
 \label{eq:near-critical-separation}
\end{equation}
Indeed, the near-throat part of \eqref{eq:separation} is equivalently $2\int_0^{\ell_0}d\ell/\sqrt{r_n^2-J^2+(r_n/n)\ell^2}$ up to terms finite in this limit, where $\ell_0>0$ is a fixed small distance from the throat. Inverting the logarithm and substituting into \eqref{eq:explicit-kernel} yields
\begin{equation}
 \mathcal I_n^{\mathbb H}(\Delta)
 =C_{n,d}\,e^{-[d-2+\kappa_n]\Delta}[1+o(1)],
 \qquad \Delta\to\infty,
 \label{eq:kernel-large-separation}
\end{equation}
where $C_{n,d}>0$ is independent of $\Delta$ and $o(1)$ vanishes in this limit. Multiplying by the endpoint shell volume $\Omega_{d-2}\sinh^{d-2}\!\Delta\,d\Delta$ leaves an $e^{-\kappa_n\Delta}$ tail in the separation distribution. Since $\kappa_n$ decreases with $n$, replica backreaction makes this tail longer ranged. At $n=1$, $\kappa_1=1$ and the density decays as $e^{-(d-1)\Delta}$, in agreement with \eqref{eq:vacuum-separation}. For $d=2$, $\kappa_n=1/n$ reproduces the exact BTZ kernel \eqref{eq:btz-kernel}. For $d>2$, \eqref{eq:kernel-large-separation} gives a geometric prediction for the cross-copy distribution beyond its total entropy.

\section{Field-theory reconstruction of the two-point PEE}
\label{app:field-theory-pee}

As discussed in the main text, reconstructing the PEE structure from the boundary theory would provide a first step toward a boundary origin for the emergent bulk geometry and tensor network. Appendix~\ref{app:boundary-kernel} obtains this structure from the replica geodesic graph. Here we give an independent boundary derivation for a single interval in the vacuum of CFT$_2$. The CHM map and uniformization allow the additive linear combination (ALC) proposal~\cite{Wen:2018whg,Wen:2019iyq} to reconstruct the cross-copy two-point PEE on $A\times A^*$ at general $n>0$, reproducing \eqref{eq:btz-kernel} and \eqref{eq:interval-pre}. We briefly discuss the obstacles to extending this derivation to higher dimensions.

The ALC proposal applies to systems with one spatial dimension and a definite spatial ordering, such as an infinite chain or a circle, and to highly symmetric higher-dimensional systems that effectively reduce to one spatial dimension, for general theories. In this appendix we restrict to one-dimensional spatial systems. There, the ALC is the most natural way to define the spatial distribution of a quantum-information quantity, as in the entanglement contour and entanglement negativity~\cite{Kudler-Flam:2019oru,Kudler-Flam:2019nhr,MacCormack:2020auw,Wen:2020ech,Ageev:2021ipd,Zambotti:2026vpm}. More explicitly, for a one-dimensional system with a definite spatial ordering and an ordered decomposition of a connected interval $B=B_1\cup B_2\cup B_3$, the ALC proposal gives~\cite{Wen:2018whg,Wen:2019iyq}
\begin{equation}
	\begin{aligned}
	\mathcal I(B_2,\bar B)&=\frac12\bigl[S(B_1\cup B_2)+S(B_2\cup B_3)\\
	&\qquad-S(B_1)-S(B_3)\bigr],
	\end{aligned}
	\label{eq:field-theory-alc}
\end{equation}
Here \(\bar B\) is the complement of $B$ in the full pure system, and $\mathcal I(B_2,\bar B)=s_B(B_2)$ denotes the contribution of $B_2$ to the entanglement entropy $S(B)$. For this ordered decomposition, the ALC expression equals half the conditional mutual information~\cite{Rolph:2021nan}
\begin{equation}
	\mathcal I(B_2,\bar B)
	=\frac12 I(B_2:\bar B|B_1).
	\label{eq:field-theory-cmi}
\end{equation}
Applying the ALC expression to infinitesimal ordered intervals gives the two-point density as a mixed endpoint derivative of connected-interval entropies. In particular, for the vacuum on a real axis with coordinate $z$, the interval entropy~\cite{Calabrese:2004eu} and its corresponding two-point PEE are
\begin{equation}
	\begin{aligned}
		S_{\rm vac}([z_1,z_2])
		&=\frac c3\log\frac{z_2-z_1}{\sqrt{\epsilon_{z_1}\epsilon_{z_2}}}+C_0,\\
		\mathcal I_{\rm vac}(z_1,z_2)
		&=\frac12\partial_{z_1}\partial_{z_2}S_{\rm vac}([z_1,z_2])\\
		&=\frac{c}{6(z_1-z_2)^2},\qquad z_1<z_2.
	\end{aligned}
	\label{eq:field-theory-vacuum-density}
\end{equation}
Here $\epsilon_{z_i}$ are local endpoint cutoffs and $C_0$ is an endpoint-independent constant. The mixed derivative removes both $C_0$ and terms depending on only one endpoint. This reconstruction uses connected-interval entropies and does not impose the normalization condition on arbitrary disconnected regions.

We now apply the ALC proposal to the canonical purification $|\Psi_n\rangle$ of the normalized power state for the vacuum interval $A=(-R,R)$. For any subregion $B\subset A\cup A^*$, write
\begin{equation}
\begin{aligned}
 \sigma_{B,n}&=\Tr_{(A\cup A^*)\setminus B}|\Psi_n\rangle\langle\Psi_n|,\\
 S_{\Psi_n}(B)&=-\Tr(\sigma_{B,n}\log\sigma_{B,n}).
\end{aligned}
 \label{eq:field-theory-subregion-state}
\end{equation}
The entropies entering the ALC proposal are these von Neumann entropies at fixed $n$, with $S_{\Psi_n}(A)=\wtS_n(A)$. For a proper subregion $B\subset A$, $\sigma_{B,n}$ generally differs from $\rho_B^n/\Tr\rho_B^n$, since taking a normalized power and tracing out $A\setminus B$ do not commute.

The boundary CHM map gives~\cite{Casini:2011kv}
\begin{equation}
 U_A\rho_{A,n}U_A^\dagger
 =\frac{e^{-b_n\H}}{\mathcal Z(b_n)},
 \qquad b_n=2\pi n,
 \label{eq:field-theory-chm-power}
\end{equation}
where $\H$ generates dimensionless hyperbolic time, $b_n=\beta_n/R$ is the dimensionless inverse temperature, and $\beta_n=2\pi nR$ is the physical inverse temperature. Here $R$ is retained, unlike the convenient choice $R=1$ in Appendix~\ref{app:chm-purification}. As shown in Appendix~\ref{app:chm-purification}, $U_A\otimes U_A^*$ maps $|\Psi_n\rangle$ to the thermofield double on $\mathbb H_L^1\cup\mathbb H_R^1$, with $A$ mapped to $L$ and $A^*$ to $R$.

To apply the ALC proposal, we must fix a common spatial ordering of the two copies. In the reflected geometry, corresponding endpoints of $A$ and $A^*$ are joined, giving the cyclic order $x:-R\to R$ along $A$, followed by $y:R\to -R$ along $A^*$. The inverse spatial CHM map,
\begin{equation}
\begin{aligned}
 x&=R\tanh\frac{q_L}{2},\qquad y=R\tanh\frac{q_R}{2},\\
 q(s)&=\log\frac{R+s}{R-s},\qquad s=x,y,
\end{aligned}
 \label{eq:field-theory-copy-ordering}
\end{equation}
transfers this order to the hyperbolic boundaries, where $q_L$ increases along $L$ and $q_R$ decreases along $R$. Both copies use the same coordinate function $q(s)$, with opposite orientations in the combined system.

This ordering is explicit in the Euclidean preparation of the thermofield double. The dimensionless strip $0\leq\tau\leq\pi n$, with $w=q+i\tau$, is mapped to the upper half-plane by
\begin{equation}
 z=e^{w/n},\qquad
 z_L=e^{q_L/n},\qquad z_R=-e^{q_R/n}.
 \label{eq:field-theory-uniformization}
\end{equation}
The two copies become the positive and negative real axes, respectively, with their matching ends joined at $z=0$ and at infinity. The resulting compactified real axis has precisely the cyclic order inherited from the doubled interval.

Let $B(q_L,q_R)$ consist of the half-lines $(-\infty,q_L]$ on $L$ and $(-\infty,q_R]$ on $R$. Their common end at $q=-\infty$ maps to $z=0$, so together they form the connected interval $[-e^{q_R/n},e^{q_L/n}]$. For local hyperbolic cutoffs $\delta_L$ and $\delta_R$, the real-axis cutoffs are $\epsilon_{z_i}=|dz_i/dq_i|\delta_i$. The vacuum interval entropy in \eqref{eq:field-theory-vacuum-density} therefore gives
\begin{equation}
\begin{aligned}
 &S_{\Psi_n}(B(q_L,q_R))\\
 &\quad=\frac c3\log\!\left[
 \frac{2n}{\sqrt{\delta_L\delta_R}}
 \cosh\frac{q_L-q_R}{2n}\right]+C_0.
\end{aligned}
 \label{eq:field-theory-cross-interval-entropy}
\end{equation}
Applying the ALC proposal to infinitesimal intervals at its two endpoints yields, with $\Delta=|q_L-q_R|$,
\begin{equation}
 \begin{aligned}
 \mathcal I_n^{\mathbb H}(\Delta)
 &=-\frac12\partial_{q_L}\partial_{q_R}
 S_{\Psi_n}(B(q_L,q_R))\\
 &=\frac{c}{24n^2}\operatorname{sech}^2\!\left(\frac{q_L-q_R}{2n}\right).
 \end{aligned}
 \label{eq:field-theory-hyperbolic-density}
\end{equation}
The minus sign follows from the opposite orientations, since increasing $q_R$ moves the negative real-axis endpoint to the left. The mixed derivative removes all endpoint-local cutoff terms and $C_0$. This directly reconstructs \eqref{eq:btz-kernel} from connected-interval entropies within the ALC proposal.

Finally, the two-point PEE transforms as an endpoint measure under the inverse CHM map,
\begin{equation}
\begin{aligned}
 \mathcal I_n(x,y)&=q'(x)q'(y)\,
 \mathcal I_n^{\mathbb H}(|q(x)-q(y)|),\\
 q'(s)&=\frac{2R}{R^2-s^2},\qquad s=x,y.
\end{aligned}
 \label{eq:field-theory-copy-density}
\end{equation}
This transformation follows from the mixed entropy derivative, with each endpoint contributing its Jacobian. Substitution reproduces \eqref{eq:interval-pre}, establishing agreement with the replica geodesic graph. As checks, the density satisfies
\begin{equation}
 \begin{aligned}
 \int_{-\infty}^{\infty}dq_R\,
 \mathcal I_n^{\mathbb H}(|q_L-q_R|)&=\frac{c}{6n},\\
 \mathcal I_1(x,y)&=\frac{cR^2}{6(R^2-xy)^2}.
 \end{aligned}
 \label{eq:field-theory-density-checks}
\end{equation}
The first identity gives \eqref{eq:interval-copy-marginal} upon multiplication by $q'(x)$. The second is the $d=2$ vacuum kernel \eqref{eq:vacuum-copy-kernel}. This boundary derivation applies to the single-interval continuum limit of CFT$_2$ at any $n>0$ within the ALC proposal, without a large-$c$ assumption.

Extending this derivation to higher-dimensional spherical regions is nontrivial. First, the ALC reconstruction relies on a one-dimensional spatial ordering and interval endpoints. Spherical symmetry alone does not provide an analogous prescription for resolving arbitrary pairs of boundary points. Second, the strip uniformization \eqref{eq:field-theory-uniformization} used to obtain the subregion entropies is specific to two dimensions. In higher dimensions, varying $n$ changes the temperature relative to the hyperbolic curvature scale, so these entropies cannot be obtained by the same conformal map. We leave the field-theory reconstruction of the higher-dimensional two-point PEE to future work.

\end{document}